# An Autoignition Study of *iso*-Butanol: Experiments and Modeling

B.W. Weber[1]    S.S. Merchant[2]     C.J. Sung[1]    W.H. Green[2]

[1]*Department of Mechanical Engineering, University of Connecticut, Storrs, CT, 06029*
[2]*Department of Chemical Engineering, Massachusetts Institute of Technology, Cambridge, MA, 02139*

The autoignition delays of *iso*-butanol, oxygen, and nitrogen mixtures have been measured in a heated rapid compression machine (RCM). At compressed pressures of 15 and 30 bar, over the temperature range 800-950 K, and for equivalence ratio of $\phi = 0.5$ in air, no evidence of an NTC region of overall ignition delay is found. By comparing the data from this study taken at $\phi = 0.5$ to previous data collected at $\phi = 1.0$ (Weber et al. 2013), it was found that the $\phi = 0.5$ mixture was less reactive (as measured by the inverse of the ignition delay) than the $\phi = 1.0$ mixture for the same compressed pressure. Furthermore, a recent chemical kinetic model of *iso*-butanol combustion was updated using the automated software Reaction Mechanism Generator (RMG) to include low-temperature chain branching pathways. Comparison of the ignition delays with the updated model showed reasonable agreement for most of the experimental conditions. Nevertheless, further work is needed to fully understand the low temperature pathways that control *iso*-butanol autoignition in the RCM.

## 1. Introduction

The demand for a clean, renewable biofuel increases as new benchmarks are legislated amid increased pressure to reduce the world's dependence on fossil fuels for energy and chemicals. Biobutanol is considered an advanced biofuel – superior to ethanol in terms of higher energy density, lower vapor pressure, and lower hygroscopicity (Nigam et al. 2011), with several practical positive effects on combustion engines (Dernotte et al. 2009). Because of the potential advantages of butanol over current generation biofuels, companies have begun to commercialize butanol from biological sources. In particular one butanol isomer – *iso*-butanol – has gained popularity because of its high octane rating and ease of industrial scale production, as demonstrated by Gevo Inc. and others (Yanowitz et al. 2011; Smith et al. 2010).

Because of the recent interest in determining the combustion properties of butanol, many kinetic models have been constructed, for example by Sarathy et al. (2012) and Hansen et al. (2012). Substantial progress has been made in the last few years to improve the ability of models to predict combustion phenomena at extreme conditions, improving predictions of low temperature, high pressure ignition delays, for example. These types of extreme conditions are important for models to faithfully predict because they are the ranges in which new advanced engine concepts will operate. Nevertheless, despite the rapid improvement of the modeling of butanol combustion, the ability of existing models to predict new experimental data is often lacking. Existing models often struggle to predict *a priori* the autoignition of the butanol isomers under off-stoichiometric conditions at high pressure and low temperature. Further work is still needed to develop truly predictive models.

In this work, new ignition delay measurements of the autoignition of *iso*-butanol acquired in a heated Rapid Compression Machine (RCM) are presented. The conditions presented in this work are selected to complement previous studies in the RCM. In particular, conditions at 15 and 30 bar pressure for equivalence ratio of $\phi = 0.5$ are presented and compared to data previously collected at $\phi = 1.0$ by Weber et al. (2013). In addition, a model for *iso*-butanol combustion is built using the open-source software, Reaction Mechanism Generator (RMG). The model is compared with the new and existing RCM data and discussion of the important pathways of *iso*-butanol decomposition is presented.

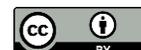

## 2. Methods

2.1 Experimental Methods

The experimental facility consists of a rapid compression machine, a fuel mixture preparation facility, and diagnostics. For mixture preparation, the fuel and oxidizer pre-mixtures are prepared in a stainless steel mixing tank. The volume of the tank is approximately 17 L so that multiple experiments can be run from a single batch. The liquid fuel (*iso*-butanol, 99.99% purity) is massed to a precision of 0.01 g in a syringe before being injected into the mixing tank through a septum. The proportions of oxygen (99.9999% purity) and nitrogen (99.9995% purity) are determined by specifying the oxidizer composition (for these experiments, the ratio of oxygen to nitrogen is fixed to that of air), the equivalence ratio, and the total mass of fuel. The gases are added to the mixing tank manometrically at room temperature. The mixture is stirred by a magnetic vane. The mixing tank, reaction chamber, and all lines connecting them are equipped with heaters to prevent condensation of the fuel. After filling the tank, the heaters are turned on and the system is allowed approximately 1.5 hours to equilibrate. This procedure has been validated previously in studies by Weber et al. (2011), Kumar et al. (2009), and Das et al. (2012). In these studies, the concentration of n-butanol, n-decane, and water were verified by GCMS, GC-FID, and GC-TCD, respectively.

The RCM used for these experiments is a pneumatically-driven/hydraulically-stopped arrangement. At the start of an experimental run the piston rod is held in the retracted position by hydraulic pressure while the reaction chamber is vacuumed to less than 1 Torr. Then, the reaction chamber is filled with the required initial pressure of the test gas mixture from the mixing tank. The compression is triggered by releasing the hydraulic pressure. The piston assembly is driven forward to compress the test mixture by high pressure nitrogen. The gases in the test section are brought to the compressed pressure ($P_C$) and compressed temperature ($T_C$) conditions in approximately 30–40 milliseconds. The piston in the reaction chamber is machined with specifically designed crevices to ensure that the roll-up vortex effect is suppressed and homogeneous conditions in the reaction chamber are promoted. In the present operation procedure $P_C$ and $T_C$ can be varied independently by adjusting the Top Dead Center (TDC) piston clearance, the stroke of the piston, the initial temperature ($T_0$), and the initial pressure ($P_0$) of the test charge. The pressure in the reaction chamber is monitored during and after compression by a Kistler Type 6125B dynamic pressure transducer. During the filling of the mixing tank and reaction chamber prior to compression, the pressure is monitored by an Omega Engineering PX-303 static pressure transducer.

Figure 1 shows a representative pressure trace from an experiment using *iso*-butanol in the RCM. The definition of the end of compression and the ignition delays are indicated on the figure. The end of compression time is defined as the time when the pressure reaches its maximum before the ignition occurs. The point of ignition is defined as the maximum of the time derivative of the pressure, in the time after the end of compression. The ignition delay is then the time difference between the point of ignition and the end of compression.

Due to heat loss from the test mixture to the cold reactor walls, the pressure and temperature will drop after the end of compression. To properly account for this effect in numerical simulations, a non-reactive pressure trace is taken that corresponds to each unique $P_C$ and $T_C$ condition studied. The non-reactive pressure trace is acquired by replacing the oxygen in the oxidizer with nitrogen, so that a similar specific heat ratio is maintained, but the heat release due to exothermic oxidation reactions is eliminated. A representative non-reactive pressure trace is also shown in Figure 1. This non-reactive pressure trace is converted to a volume trace for use in simulations in CHEMKIN-Pro (2011) using the temperature dependent specific heat ratio.

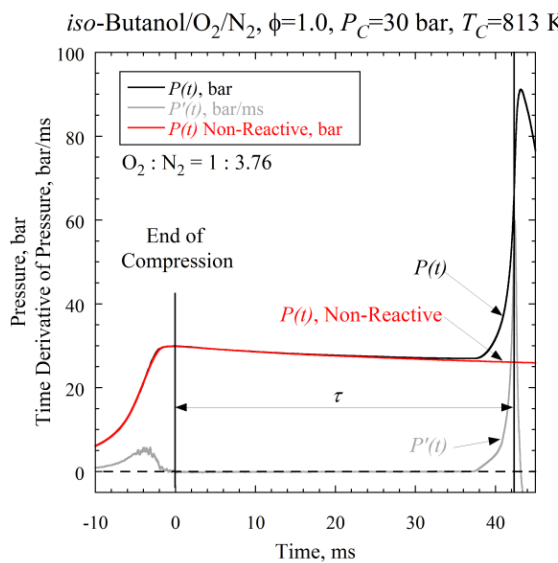

**Figure 1:** Representative pressure trace indicating the definition of the ignition delay and the corresponding non-reactive pressure trace.

Each unique $P_C$ and $T_C$ condition is repeated at least 6 times to ensure repeatability of the experiments. The experiment closest to the mean of the runs at a particular condition is chosen for analysis and presentation. The standard deviation of all of the runs at a condition is less than 10% of the mean in all cases. Furthermore, to ensure reproducibility, each new mixture preparation is checked against a previous experiment.

2.2 Numerical Methods

Simulations are performed using the Closed Batch Homogeneous Reactor model in CHEMKIN-Pro. The reactor volume is prescribed as a function of time by the non-reactive volume trace described earlier. This type of simulation captures the heat loss effects during the compression stroke and the post-compression event, which allows the simulation to more accurately describe the actual thermodynamic conditions in the reaction chamber. It also captures the effect of any reactions that occur during compression. This type of simulation is referred to as a VPRO simulation.

VPRO simulations are used to calculate the temperature at the end of compression, $T_C$. This temperature is used as the reference temperature for reporting the ignition delay. This approach requires the assumption of an adiabatic core of gases in the reaction chamber, which is facilitated on the present RCM by the optimized creviced piston described previously. Simulations to determine $T_C$ are conducted with and without reactions in the reaction mechanism. For all the conditions investigated herein, the pressure and temperature at TDC are the same whether or not reactions are included in the simulation, indicating insignificant reactivity during the compression stroke.

2.3 Mechanism Development

The *iso*-butanol kinetic model presented in this study is created using the automatic Reaction Mechanism Generator (RMG) version 4.0 (Green et al 2013). Details of the implementation of the RMG algorithm are given in Harper et al. (2011). The mechanism produced in this work is an updated version of the mechanism presented in the work by Hansen et al. (2012) and Merchant et al. (2012) - specifically detailed low temperature peroxy pathways have been included in the current mechanism. The following classes of reactions are present in the current model:

1) Addition of O2 to parent fuel radical (R + O2 = ROO)
2) ROO isomerization to QOOH (including Waddington type of reaction)
3) ROO concerted elimination to Enol + HO2
4) ROO abstraction from RH to give ROOH and decomposition to RO + OH
5) QOOH cyclization to corresponding cyclic ether
6) Addition of O2 to QOOH (QOOH + O2 = O2QOOH)
7) O2QOOH elimination to ketohydroperoxide and HO2
8) Isomerization of O2QOOH to HO2Q'OOH
9) Decomposition HO2Q'OOH to ketohydroperoxide + OH
10) Decomposition of ketohydroperoxide to give oxygenated radical + OH

R refers to parent fuel radical such as CH3CH(CH3)CH*OH, ROO refers to alkylperoxy radical such as CH3CH(CH3)CH(OO*)OH, QOOH refers to hydroxyalkylhydroperoxide radical such as CH2*CH(CH3)CH(OOH)OH. A recently published novel reaction pathway by Welz et al. (2013) of gamma-QOOH radical (gamma refers to the third carbon from the alcohol group) to stable product + water is also included in the mechanism. As no rate coefficient was presented for the gamma-QOOH pathway, we estimate the rate constant in the current model where the Arrhenius factor is that of a typical QOOH 6-membered ring cyclization rate (Sarathy et al. 2012) and the activation energy of 14.5 kcal/mol which is the barrier height calculated by Welz et al. (2013) . As many of the rate coefficients are unavailable in literature, they are estimated using RMGs group contribution method based on alkane peroxy chemistry and therefore have significant uncertainty (factor of 5 – 10) associated with them.

3. Results and Discussion

The experimental ignition delays of *iso*-butanol measured in this study at $P_C = 15$ and 30 bar and $\phi = 0.5$ are shown in Figure 2. The error bars are equal to twice the standard deviation of all the runs at that condition. The lines are curve fits to the data. The circles represent the 15 bar data, while the squares represent the 30 bar data. Also shown in Figure 2 are the experimental ignition delays previously measured for *iso*-butanol at $\phi = 1.0$ and $P_C = 15$ and 30 bar. The $\phi = 0.5$ cases are shown in blue and the $\phi = 1.0$ cases are shown in red.

For both equivalence ratios, the 15 bar cases are less reactive than the 30 bar cases, as judged by the inverse of the ignition delay. Furthermore, in comparing the $\phi = 1.0$ data to the $\phi = 0.5$ data at the same compressed pressure, it is seen that the strong equivalence ratio dependence of the ignition delays previously measured for two other isomers of butanol, *n*-butanol (Weber et al. 2011), and *tert*-butanol (Weber et al. 2013), is also present for *iso*-butanol.

Figures 3–4 show comparisons of the experimentally measured ignition delays with the results from the updated model. The simulations shown in these figures are the VPRO type of simulation. Some simulated cases at low temperature did not ignite within the experimental duration, and hence they are not shown in the figures.

As shown in Figure 3, the model is able to predict the ignition delay of lean *iso*-butanol mixtures reasonably well, especially at the lower pressure. At the higher pressure, the ignition delay is over-predicted by approximately a factor of 2 in the lower temperature range of the data. At $\phi = 1.0$, larger discrepancy is noted in Figure 4, as the model over-predicts the data by a factor of 2~2.5, even at the lower pressure. It is interesting to note that the agreement is much better for the $\phi = 0.5, 15$ bar case compared to the $\phi = 1.0, 15$ bar case, but the agreement is similar between the equivalence ratios at 30 bar.

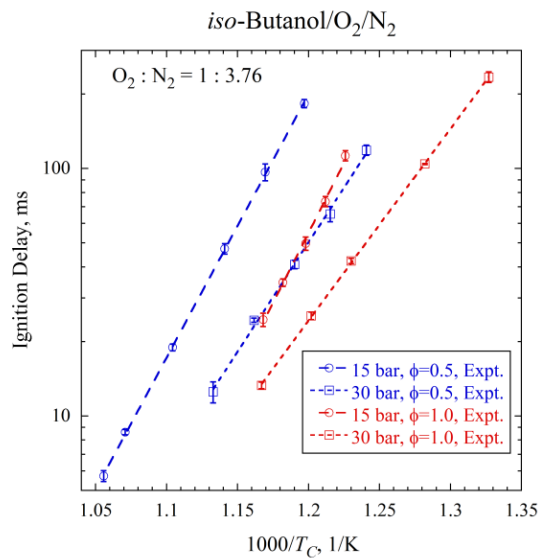

**Figure 2:** Comparison of the experimentally measured ignition delays of *iso*-butanol at two compressed pressures, $P_C = 15$ bar (circles) and $P_C = 30$ bar (squares), and two equivalence ratios, $\phi = 0.5$ (blue) and $\phi = 1.0$ (red).

## 4. Conclusions

In summary, new experimental data has been presented for the autoignition of *iso*-butanol in a heated RCM. The new data were collected at compressed pressures of 15 and 30 bar, over the temperature range 800-950 K, and for equivalence ratio $\phi = 0.5$. The oxidizer was nitrogen/oxygen air. In addition, an existing chemical kinetic model for the combustion of *iso*-butanol was updated by using the automated software RMG to include many low-temperature pathways. The newly collected experimental data was compared with existing experimental data for *iso*-butanol at $\phi = 1.0$, and the $\phi = 0.5$ mixtures were found to be less reactive than the $\phi = 1.0$ mixtures at the same compressed pressure. Further comparison of all of the experimental data with the newly updated model showed reasonable agreement (within a factor of 2) across most conditions. Due to the large uncertainties associated with many of the newly added reactions, work is ongoing to improve the model predictions.

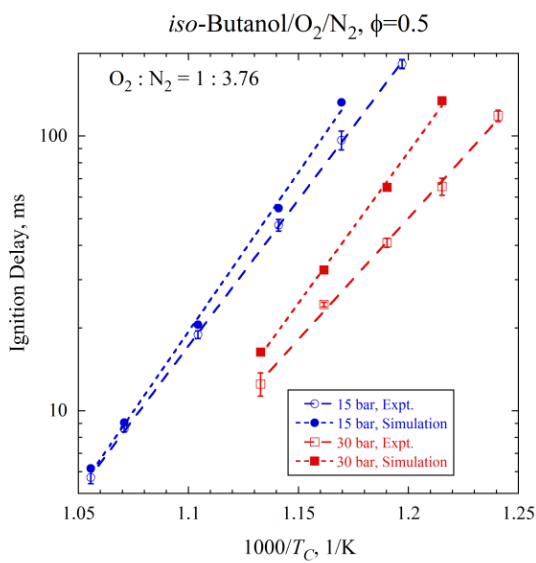

**Figure 3:** Comparison of experimentally measured ignition delays at $\phi = 0.5$ with the model. Open symbols are experiments and filled symbols are simulations. Blue is 15 bar and red is 30 bar.

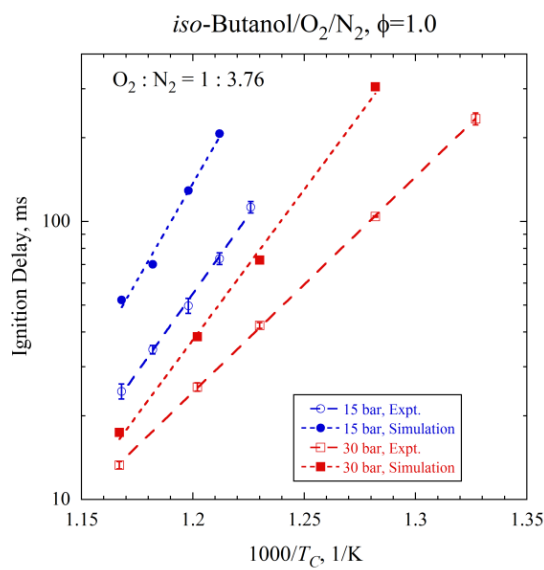

**Figure 4:** Comparison of experimentally measured ignition delays at $\phi = 1.0$ with the model. Open symbols are experiments and filled symbols are simulations. Blue is 15 bar and red is 30 bar.

**Acknowledgements**

This work was supported as part of the Combustion Energy Frontier Research Center, an Energy Frontier Research Center funded by the US Department of Energy, Office of Science, Office of Basic Energy Sciences, under Award Number DE-SC0001198.

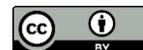

Yanowitz, Christensen, Mccormick, Yanowitz, Christensen, and Mccormick. 2011. Utilization of Renewable Oxygenates as Gasoline Blending Components Utilization of Renewable Oxygenates as Gasoline Blending Components. NREL Report TP-5400-50791.

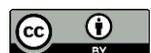